\documentclass[twocolumn,showpacs,amssymb,floatfix]{revtex4}
\usepackage{graphicx}
\usepackage{dcolumn}
\usepackage{bm}
\begin{document}
\title{Bound states in the continuum in open Aharonov-Bohm rings}
\author{Evgeny N. Bulgakov$^1$, Konstantin N. Pichugin$^1$, Almas F. Sadreev$^1$,
and Ingrid Rotter$^3$}
\address{$^1$ Institute of Physics, Academy of Sciences, 660036 Krasnoyarsk,
Russia}
\address {$^3$ Max-Planck-Institut f\"ur Physik
komplexer Systeme, D-01187 Dresden, Germany }
\date{\today}
\begin{abstract}
Using formalism of effective Hamiltonian we consider bound states
in continuum (BIC). They are those eigen states of non-hermitian
effective Hamiltonian which have real eigen values. It is shown
that BICs are orthogonal to open channels of the leads, i.e.
disconnected from the continuum. As a result BICs can be
superposed to transport solution with arbitrary coefficient and
exist in propagation band. The one-dimensional Aharonov-Bohm rings
that are opened by attaching single-channel leads to them allow
exact consideration of BICs. BICs occur at discrete values of
energy and magnetic flux however it's realization strongly depend
on a way to the BIC's point.
\end{abstract}
\pacs{03.65.Ge, 03.65.Nk, 05.60.Gg, 72.20.Dp} \maketitle

\section{Introduction.}
%%%%%%%%%%%%%%%%%%%%%%%%%%%%%%%%%%%%%%%%%%%%%%%%%%%%%%%%%%%%%%%%%
In 1929,  von Neumann and Wigner \cite{neumann} firstly pointed to
the existence of discrete solutions of the single-particle
Schr\"odinger equation embedded in the continuum of positive
energy states. Their analysis has been examined by Stillinger and
Herrick \cite{stillinger} in the context of possible bound states
(BICs) in atoms and molecules. It has been demonstrated by Newton
\cite{newton} that strong coupling between scattering channels can
give rise to BIC. BIC can be observed in the stationary
transmission as resonant states with width which tends to zero as
at least two physical parameters vary continuously as it was
formulated by Friedrich and Wintgen \cite{friedrich}, who have
also given the example of the hydrogen atom in a magnetic field.
Such a BIC is a very fragile structure. A small perturbation
transforms it into narrow resonance. Nevertheless, Capasso {\it et
al} \cite{capasso} have reported direct evidence for BIC's in a
semiconductor superlattice.

For better understanding of the phenomenon of BIC's in transport
through electronic devices it is useful to study as simple quantum
system as possible. Robnik \cite{robnik} has shown that a simple
separable two-dimensional Hamiltonian can develop BIC under
perturbation of open channels. An explicit proof of an existence
of BIC's was presented recently by Cederbaum {\it et al}
\cite{cederbaum} in the molecular system, if the electronic and
the nuclear motions are coupled. In the present letter we consider
the open Aharonov-Bohm (AB) rings which are good candidates to
observe BICs for the external magnetic field and energy of
incident electron can be easily varied experimentally. Moreover
the one-dimensional AB rings allow to treat BICs wholly
analytically. A phenomenon of zero resonance widths at discrete
values of energy of incident particle and some relevant physical
parameter was established in many works since the work by
Shahbazyan and M.E. Raikh
\cite{shahbazyan,olendski,guevara,rotter1,sadreev1,sadreev2,rowe,na,sadreev3},
among of them external magnetic flux was considered in
\cite{wunsch,orellana}. In this letter we focus on the scattering
wave function in the vicinity and at BIC's point and how BIC
participated in transport.
%%%%%%%%%%%%%%%%%%%%%%%%%%%%%%%%%%%%%%%%%%%%%%%%%%%%%%%%%%%%%
\section{The one-dimensional ring}
\begin{figure}[ht]
\includegraphics[width=.35\textwidth,height=0.23\textheight]{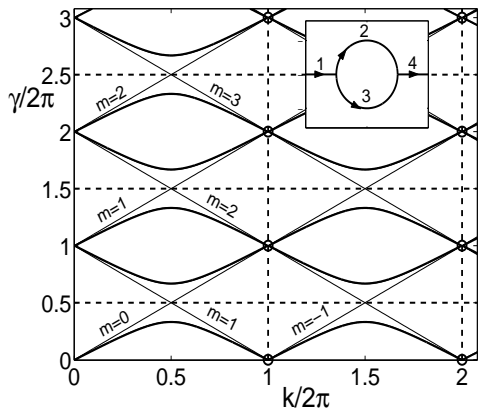}
\caption{Transmission zeros $|t|^2=0$ and ones $|t|^2=1$ of the
one-dimensional ring as function of the wave number $k$ and flux
$\gamma$. The zeros (ones) are shown by dashed (solid) lines. The
thin solid lines represent the eigenenergies of the closed ring.}
\label{fig1}
\end{figure}

%%%%%%%%%%%%%%%%%%%%%%%%%%%%%%%%%%%%%%%%%%%%%%%%%%%%%%%%%%%%%%%
Following Xia \cite{xia} we write the wave functions in the
segments of the structure shown in inset of Fig. \ref{fig1} as
%------------------------------------------------------------------(1)
\begin{equation}\label{psi1d}
\begin{array}{rcl}
\psi_1(x)&=&\exp(ikx)+r\exp(-ikx),\\
\psi_2(x)&=&a_1\exp(ik^-x)+a_2\exp(-ik^+x),\\
\psi_3(x)&=&b_1\exp(ik^+x)+b_2\exp(-ik^-x),\\
\psi_4(x)&=&t\exp(ikx),
\end{array}
\end{equation}
where $k^-=k-\gamma, ~k^+=k+\gamma, ~\gamma=2\pi\Phi/\Phi_0,
~\Phi=B\pi R^2$ is the magnetic flux, $ \Phi_0=2\pi\hbar c/e$. The
ring length $2\pi R$ is chosen as unit. The boundary conditions
(the continuity of the wave functions and the conservation of the
current density) allow to find all coefficients in (\ref{psi1d}).
We write the corresponding equation in matrix form
%---------------------------------------------------------------(2)
\begin{equation}\label{AS}
\hat{F}\overrightarrow{\psi}= \overrightarrow{g},
\end{equation}
where $\hat{F}(k,\gamma)$ is the following matrix
%--------------------------------------------------------------(3)
\begin{equation}\label{matrixF}
  \left(\begin{array}{cccccc}
    -1 & 0 & 1 & 1 & 0 & 0 \cr
    -1 & 0 & 0 & 0 & 1 & 1 \cr
    0 & -1 & e^{ik^-/2} & e^{-ik^+/2} & 0 & 0 \cr
    0 & -1 & 0 & 0 & e^{ik^+/2} & e^{-ik^-/2} \cr
    1 & 0 & \frac{k^-}{k}& -\frac{k^+}{k}& \frac{k^+}{k} &
    -\frac{k^-}{k}\cr
    0 & -1 &\frac{k^-}{k}e^{i\frac{k^-}{2}} & -\frac{k^+}{k}e^{-i\frac{k^+}{2}} &
    \frac{k^+}{k}e^{i\frac{k^+}{2}} &
    -\frac{k^-}{k}e^{-i\frac{k^-}{2}} \end{array}\right),
\end{equation}
$\overrightarrow{g}^{T}=(1~ 1 ~0 ~0 ~1 ~0)$. The vector
$\overrightarrow{\psi}^{T}=(r ~t ~a_1 ~a_2 ~b_1 ~b_2)$ is the
solution for the scattering wave function:
%---------------------------------------------------------------------(4)
\begin{equation}\label{S}
\begin{array}{rcl}
r&=&2(3\cos{k}-4\cos{\gamma}+1)/Z,\\
t&=&16i(\sin{\frac{k}{2}}\cos{\frac{\gamma}{2}})/Z\\
a_1&=&2(2e^{i\gamma}-3e^{-ik}+1)/Z,\\
a_2&=&2(e^{ik}+1-2e^{i\gamma})/Z,\\
Z&=&8\cos{\gamma}-9e^{-ik}-e^{ik}+2,\\
\end{array}
\end{equation}
$b_{1,2}(k,\gamma)=a_{1,2}(k,-\gamma)$. In Fig. \ref{fig1} we show
lines of the transmission zeros ($|t(k,\gamma)|=0$, dashed lines)
which cross the lines of the transmission ones ($|t(k,\gamma)|=1$,
solid lines) at points
%-----------------------------------------------------------------(5)
\begin{equation}\label{BIC points}
\begin{array}{rcl}
  k_m&=&2\pi m, m=\pm 1, \pm2, \ldots,\\
  \gamma_n&=&2\pi n, n=0, \pm 1, \pm2, \ldots.
\end{array}
\end{equation}
As can be seen from the expression for the denominator $Z$ in Eqs.
(\ref{S}), the imaginary part of the  poles vanishes at these
points. Simultaneously, there is a degeneracy of eigenenergies of
closed ring $(k_m-\gamma)^2$. Here $m$ is the magnetic quantum
number that defines the eigen functions of the closed ring
$\psi_m(x)=\exp(ik_m x)$. The point $k=0$ is excluded from the
consideration since it gives zero conductance. The peculiar points
(\ref{BIC points}) were shown in \cite{wunsch} for the case of
single lead attached to the 1d ring. To show that the bound states
in continuum (BICs) appear at the points (\ref{BIC points}), let
us consider one of the points, say, ${\bf s}_0=
(k_1,\gamma_1)=2\pi(1,~1)$. All the other points are equivalent
because of the periodical dependence of the system on $k$ and
$\gamma$. In the vicinity of the point ${\bf s}_0$ we write Eqs
(\ref{S})   in the following approximated form
%------------------------------------------------------------------(6)
\begin{eqnarray}\label{Sappr}
%\begin{array}{c}
t\approx \frac{\Delta k}{\Delta k+i(\Delta\gamma)^2/2},~ r\approx
\frac{i(3\Delta k^2-4\Delta\gamma^2)}{4(2\Delta
k+i\Delta\gamma^2)},\nonumber\\
 a_1\approx
\frac{3\Delta k+2\Delta\gamma}{4\Delta k+2i\Delta\gamma^2},~
a_2\approx \frac{\Delta k-2\Delta \gamma}{4\Delta
k+2i\Delta\gamma^2},
%\end{array}
\end{eqnarray}
where $\Delta k=k-k_1, ~\Delta\gamma=\gamma-\gamma_1$. The
transmission amplitude in the vicinity of the point ${\bf s}_0$ in
(\ref{Sappr}) is similar to the expressions obtained for a shifted
von Neumann-Wigner potential \cite{pursey}. One can see that all
amplitudes $a_{1,2}, ~b_{1,2}$ of the inner wave functions are
singular at the point ${\bf s}_0$. Such a result for the BIC
points was firstly found by Pursey and Weber \cite{pursey}.

Eqs ({\ref{AS}) and (\ref{matrixF}) allow to show that the point
${\bf s}_0$ corresponds to the BIC one in an open one-dimensional
ring. At this point the matrix (\ref{matrixF}) takes the following
form
%--------------------------------------------------------------------(7)
\begin{equation}\label{matrixa}
  \hat{F}({\bf s}_0)=\left(\begin{array}{cccccc}
    -1 & 0 & 1 & 1 & 0 & 0 \cr
    -1 & 0 & 0 & 0 & 1 & 1 \cr
    0 & -1 & 1 & 1 & 0 & 0 \cr
    0 & -1 & 0 & 0 & 1 & 1 \cr
    1 & 0 & 0& -2 & 2 & 0 \cr
   0 & -1 & 0 & -2 & 2 & 0 \end{array}\right).
\end{equation}
The determinant of the matrix $\hat{F}({\bf s}_0)$ equals zero.
Therefore, $\hat{F}\overrightarrow{f_0}=0$. By direct substitution
of the  vector
 $\overrightarrow{f_0}^{T}=\frac{1}{2}(0 ~0 ~1 ~-1 ~-1 ~1)$
one can verify that $\overrightarrow{f}_0$ is the right
eigenvector which is the null vector. The corresponding left null
eigenvector is $\tilde{\overrightarrow{f}}_0=\frac{1}{2}(-1 ~1 ~1
~-1 ~0 ~0)$. It is well known from linear algebra, that if the
determinant of matrix $\hat{F}$ is equaled to zero, then the
necessary and sufficient condition for existence of solution of
the equation (\ref{AS}) is that the vector
$\tilde{\overrightarrow{f}}_0$ is orthogonal to vector
$\overrightarrow{g}$ \cite{smirnov}. In holds, indeed,
$\tilde{\overrightarrow{f}}_0\cdot\overrightarrow{g}=0$. The
solution of Eq. (\ref{AS}) at the point ${\bf s}_0$ can therefore
be presented as
%-----------------------------------------------------------------------(8)
\begin{equation}\label{BIC sol}
\overrightarrow{\psi}({\bf s}_0)=\alpha\overrightarrow{f}_0+
\overrightarrow{\psi_p},
\end{equation}
where $\alpha$ is an {\it arbitrary} coefficient and
$\overrightarrow{\psi_p}$ is particular transport solution of Eq.
(\ref{AS}). By direct substitution one can verify that
$\overrightarrow{\psi_p}^T=\left(0 ~1 ~\frac{3}{4} ~\frac{1}{4}
~\frac{3}{4} ~\frac{1}{4}\right)$ is the particular solution of
Eq. (\ref{AS}). It is worthwhile to note that this result
completely agrees with the scattering theory on graphs
\cite{texier1,texier2}. Texier \cite{texier1} has shown that for
certain graphs the stationary scattering state gives the solution
of the Schr\"odinger equation for the continuum spectrum apart for
discrete set of energies where some additional states are
localized in the graph and thus are not probing by scattering,
leading to the failure of the state counting method from the
scattering. Similar result was independently obtained recently by
Voo and Chu \cite{voo}.

In the vicinity of the BIC point ${\bf s}_0$ the scattering state
using (\ref{Sappr}) becomes, to leading order of $\Delta k,
~\Delta\gamma$,
%-----------------------------------------------------------------------(9)
\begin{equation}\label{psilimit}
  \overrightarrow{\psi}({\bf s})\approx\frac{\Delta\gamma\overrightarrow{f}_0+
  \Delta k\overrightarrow{\psi}_p}{\Delta
k+i\Delta\gamma^2/2},
\end{equation}
where ${\bf s}=(k,\gamma)$. Thus, the scattering state in the
nearest vicinity of the BIC point also is superposed of the BIC
vector $\overrightarrow{f}_0$ and of the particular solution
$\overrightarrow{\psi}_p$. Eq. (\ref{psilimit}) shows that the
limiting scattering wave state $\overrightarrow{\psi}$~ depends on
a way ${\bf s}\rightarrow{\bf s}_0$. If we at first take
$\Delta\gamma=0$, then obtain
$\overrightarrow{\psi}=\overrightarrow{\psi}_p$ which is a
transport solution. If we, however, choose at first $\Delta k=0$,
then have
$\overrightarrow{\psi}=\frac{2}{i\Delta\gamma}\overrightarrow{f}_0$,
i.e. the scattering state is diverging interior the ring. This
formula shows that the BIC state $\overrightarrow{f}_0$ can be
extracted from the scattering state by a special limit in
(\ref{psilimit}).
%%%%%%%%%%%%%%%%%%%%%%%%%%%%%%%%%%%%%%%%%%%%%%%%%%%%%%%%%%%%%%%%%%%
\section {Two-dimensional devices}
%%%%%%%%%%%%%%%%%%%%%%%%%%%%%%%%%%%%%%%%%%%%%%%%%%%%%%%%%%%%%%%%%%
Typical open two-dimensional structures are dots or rings with
attached leads. Numerically the transmission through them are
solving by finite-difference equations which are equivalent to the
tight-binding lattice model \cite{datta,sadreev}. The case of the
quantum dots was considered in \cite{sadreev3}. Here we present
results of computation for the two-dimensional ring with
symmetrically attached identical leads.  In Fig. \ref{fig2} we
show %, similar to Fig.\ref{fig1},
the transmission zeros (dashed lines) and the transmission ones
(solid lines) for the single-channel transmission.
\begin{figure}[ht]
\includegraphics[scale=0.4]{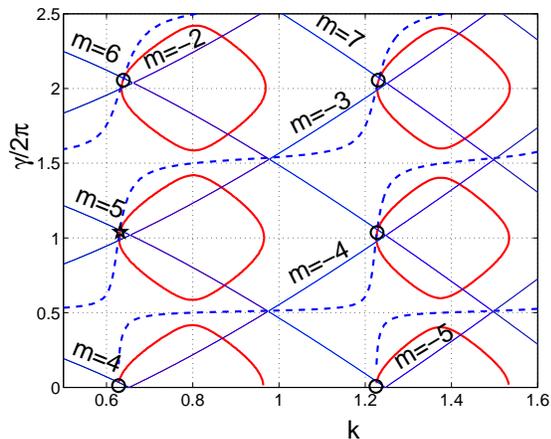}
\caption{Zeros (dashed blue lines) and ones (solid red lines) of
the transmission probability of the two-dimensional ring as
function of wave number $k=\sqrt{E-\pi^2}$ and flux $\gamma=B\pi
R^2/\Phi_0$, $\Phi_0=2\pi\hbar c/e$. $R=2.5$ is mean radius of the
ring. The width of the ring and those of the leads are equaled to
unit. The eigenenergies of the closed two-dimensional ring are
shown by thin lines. The BIC points are marked by open circles and
star.} \label{fig2}
\end{figure}

In order to find the positions and widths of the resonance states,
we explore the non hermitian effective Hamiltonian, which can be
obtained by projection of whole system onto Hilbert space of
closed system \cite{dittes,sadreev,physrep}. The effective
Hamiltonian in the basis of closed system's eigenvectors can be
written as \cite{datta,sadreev}
%-----------------------------------------------------------------------(10)
\begin{equation}
\label{Heff}
 \langle b|H_{eff}|b'\rangle=E_b\delta_{bb'}-\sum_p\sum_{C=L,R} V^C_{b,p}V^C_{b',p}e^{ik^C_p}.
\end{equation}
Here $E_b$ and $|b\rangle$ are the eigenvalues and the
eigenfunctions of closed system are given by quantum numbers $b$,
$C$ enumerates left and right leads and $p$ does the open channels
of leads. A recipe to calculate the matrix elements $V^C_{b,p}$ is
given in \cite{dittes,sadreev}. Because of energy dependence of
the effective Hamiltonian the positions and widths of the
resonance states are defined by the following nonlinear fixed
point equations \cite{physrep}
%-----------------------------------------------------------------------(11)
\begin{equation}\label{fix point}
  E_{\lambda}=Re({z_\lambda(\gamma,E_{\lambda})}), ~~ 2\Gamma_{\lambda}=
-Im({z_\lambda(\gamma,E_{\lambda})}).
\end{equation}
Here $z_{\lambda}$ are the complex eigenvalues of the effective
Hamiltonian (\ref{Heff}) $H_{eff}|\lambda)=z_{\lambda}|\lambda)$
with right eigenstates $|\lambda)$.
%\cite{physrep}.
All the points at which $\Gamma_{\lambda}=0$, i.e. the width of
the resonant transmission vanishes  are marked in Fig. \ref{fig2}
by open circles and star.
%The phenomenon of infinitely narrow
%resonances was reported for transmission through different
%structures \cite{shahbazyan,olendski,guevara,RS1,JETPLet,na}.

%\begin{figure}[ht]
%\includegraphics[width=.4\textwidth,height=0.2\textheight]{ringfig3.eps}
%width=.5\textwidth,height=0.25

The equation for the scattering wave function mapped interior the
ring $|\psi_R\rangle$ can be derived from the Lippmann-Schwinger
equation \cite{dittes,sadreev,physrep} and takes the following
form
%---------------------------------------------------------------------(12)
\begin{equation}\label{GF}
(H_{eff}-E)|\psi_R\rangle=V^L|E,L,p=1\rangle.
\end{equation}
Here $V^L$ is the coupling matrix between the left lead and the
ring provided that a particle incidents from the left lead in the
first channel. This formula is similar to (\ref{AS}) for the 1d
ring. If $Det(H_{eff}-E)\neq 0$, then in the biorthogonal basis
$|\lambda)$ the scattering wave function takes a simple form
\cite{sadreev,physrep}
%------------------------------------------------------------------------(13)
\begin{equation}
\label{psiR} |\psi_R\rangle=\sum_{\lambda}
\frac{V_{\lambda}(\gamma,E)}{E-z_{\lambda}(\gamma,E)}|\lambda),
\end{equation}
where
%-----------------------------------------------------------------------(14)
\begin{equation}\label{Wlambda}
V_{\lambda}=(\lambda|V|E,L,p=1\rangle =\int dy_B
\widetilde{\psi}_{\lambda}(y_B) \sin(\pi y_B),
\end{equation}
$\widetilde{\psi}_{\lambda}$ are the left eigen functions of
$H_{eff}$, $y_B$ runs over the boundary that connects the closed
ring and the left lead with the first channel excited ($p=1$).  We
assume that magnetic field subjects only the ring.

Let us denote a set of physical parameters of the system as ~${\bf
s}$. For example, for present case of the ring ${\bf s}=(E,
\gamma)$, although for the quantum dot ${\bf s}$ might be energy
and confined potential \cite{sadreev3}. Let us consider the point
${\bf s}_0=(E_0, \gamma_0)$ at which Eq. (\ref{fix point}) is
fulfilled $E_0=z_{\lambda_0}(E_0, \gamma_0)$ and
$\Gamma_{\lambda_0}=0$, i.e. one of the complex eigenvalues of
$H_{eff}$ is real at this point. For $E=E_0$ one have equality
$(H_{eff}-E)|\lambda_0)=0$. Comparing this equation to (\ref{GF})
we see that the eigen state $|\lambda_0)$ corresponds to the
solution of the Lippmann-Schwinger equation if there were no
ingoing current in the left lead. Respectively, the state
$|\lambda_0)$ can not give rise to outgoing currents because of
the continuity equation for the current density. In order to
fulfill that we have to consider that the eigen function
$\psi_{\lambda_0}$ does not overlap with the first channel of the
left lead, i. e. $V^L_{\lambda_0}({\bf s}_0)=0$. This may be also
established by consideration of the transmission amplitude
\cite{sadreev}
%-------------------------------------------------------------------(15)
\begin{equation}\label{trans}
t=-2\pi i\sum_{\lambda}\frac{\langle E,L|
V^L|{\lambda})({\lambda}|V^R|E,R\rangle}{E-z_{\lambda}}.
\end{equation}
 Because of symmetry of the system relative to the left and right leads
  $|V^L_{\lambda_0}|=|V^R_{\lambda_0}|$. In approaching the
point ${\bf s}\rightarrow {\bf s}_0$ the denominator
$E-z_{\lambda_0}({\bf s})\rightarrow 0$. Therefore, in order the
ratio $|V^L_{\lambda_0}({\bf s})|^2/(E-z_{\lambda_0}({\bf s}))$
remained finite in (\ref{trans}) it is necessary
$|V^L_{\lambda_0}({\bf s})|\rightarrow 0$ for ${\bf s}\rightarrow
{\bf s}_0$. Thus, at the BIC point we have orthogonality of the
righthand state ($V|E,L, p=1\rangle$) in Eq. (\ref{GF}) to the
left eigen state $(\lambda_0|$. Then, in full correspondence to
the consideration of the 1d ring (Eq. (\ref{BIC sol})), we have
the following solution for the scattering state interior the ring
%---------------------------------------------------------------------------(16)
\begin{equation}
\label{psiR0} |\psi_R({\bf s}_0)\rangle=\alpha|\lambda_0({\bf
s}_0))+|\psi_p({\bf s}_0)\rangle,
\end{equation}
where coefficient $\alpha$ is arbitrary. Right eigen function
$\psi_{\lambda_0({\bf s}_0)}$ of the effective Hamiltonian is
squared integrable and therefore is the BIC function shown in Fig.
\ref{fig3} (a). Although the BIC function is disconnected from the
the first channel of the left lead, it couples with the next
channels $p>1$ of the leads which are evanescent modes. As a
result the BIC function has exponentially small tails in the leads
as might be seen from Fig. \ref{fig3} (a).
\begin{figure}
\includegraphics[scale=0.3]{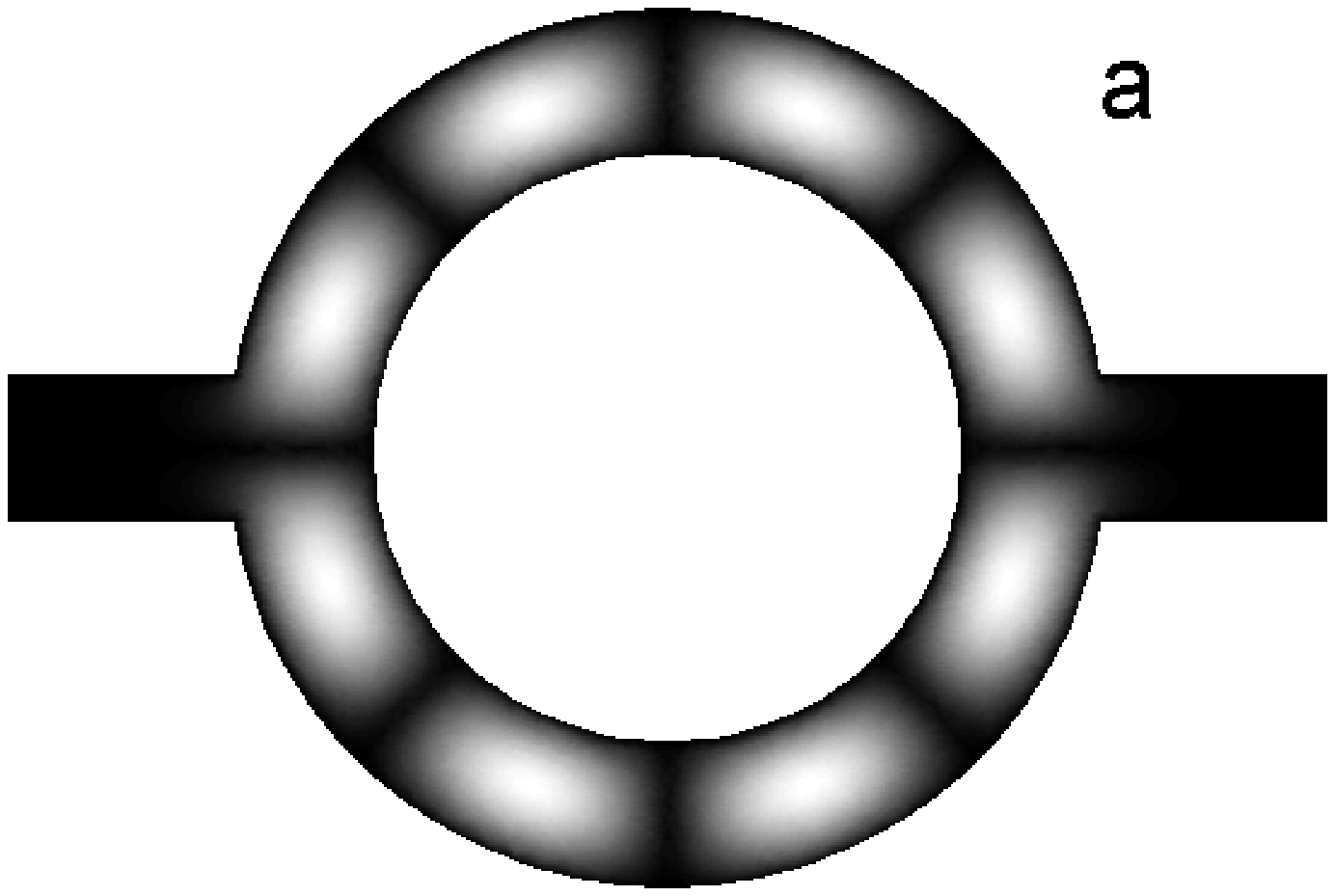}
\includegraphics[scale=0.3]{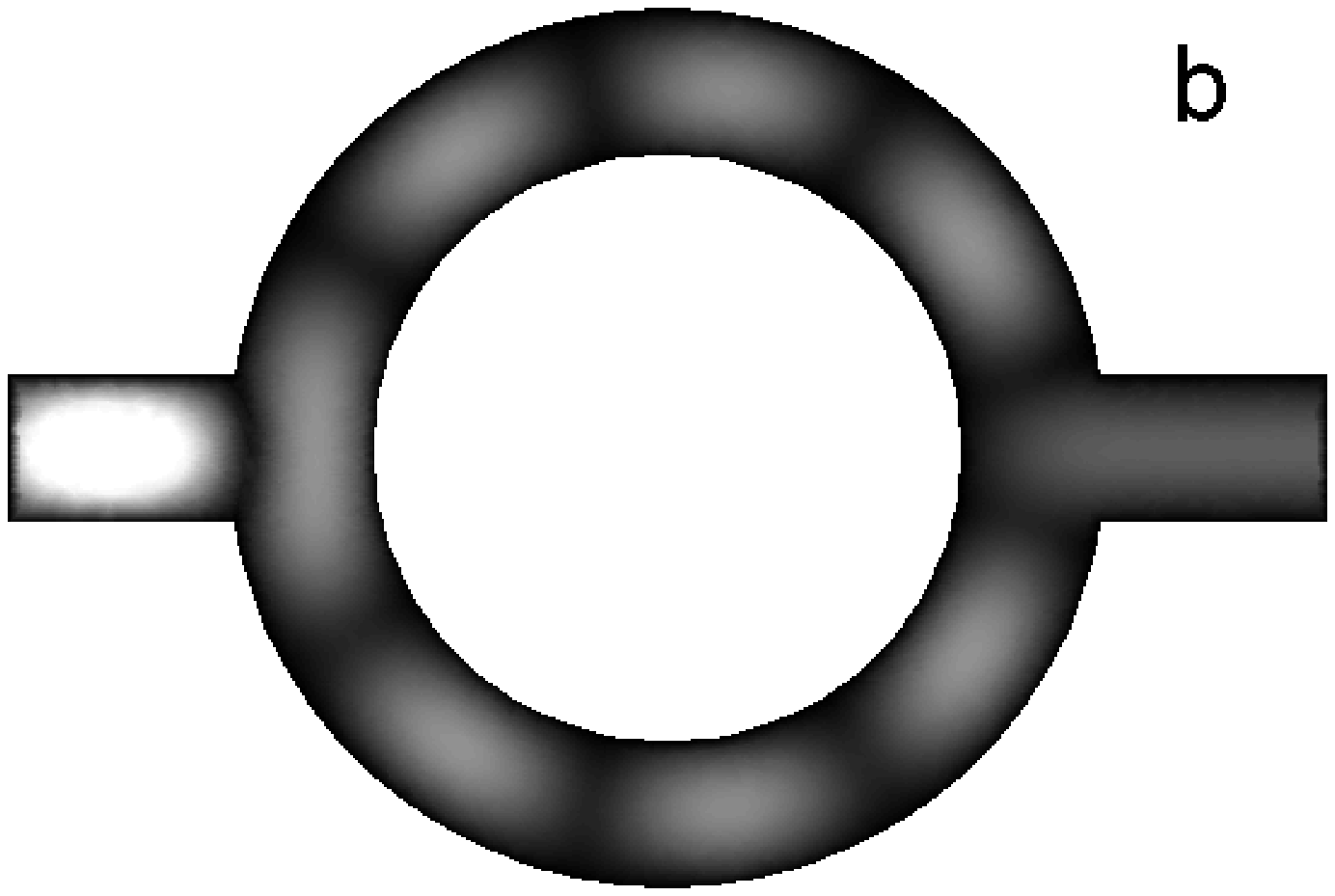}
\caption{The BIC function $|\psi_{\lambda_0}|$ which is the eigen
function of the effective Hamiltonian (\ref{Heff}) (a) and the
transport solution $|\psi_p|$ (b) at BIC point  marked in Fig.
\ref{fig2} by star.}
\label{fig3}
\end{figure}

Moreover the coupling of the 2d ring with the evanescent modes
gives rise to that as Fig. \ref{fig2} shows the BIC points are
close to but different from points at which two eigen functions of
closed 2d ring classified by magnetic quantum numbers $m$ have the
same energy. The evanescent modes have imaginary wave numbers
$k_p$ which change effectively the Hamiltonian of closed ring by
matrix $$\sum_{p\neq 1}\sum_{C=L,R}
V^C_{b,p}V^C_{b',p}e^{-|k_p|}\sim (d/R)^2$$ via Eq. (\ref{Heff}).
Therefore only for limiting case of the 1d ring $d/R\rightarrow 0$
the BIC state will consists of a pair of eigen states of closed
ring as seen from Fig. \ref{fig1} and as was confirmed by
computations.

 In the vicinity of ${\bf s}_0$  a value $E-z_{\lambda_0}(E, \gamma)$ is
small. Then we can split the summation over $\lambda$ in
(\ref{psiR}) by two parts, $\lambda=\lambda_0$ and $\lambda\neq
\lambda_0$ and similar to (\ref{BIC sol}) write the scattering
state as
%---------------------------------------------------------------------------(17)
\begin{equation}
\label{psiRR} |\psi_R({\bf s})\rangle=\alpha({\bf
s})|\lambda_0({\bf s}))+|\psi_p({\bf s})\rangle,
\end{equation}
where
%--------------------------------------------------------------------------(18)
\begin{equation}\label{ratio}
\alpha({\bf s})=\frac{V_{\lambda_0}({\bf
 s})}{E-z_{\lambda_0}({\bf s})},
\end{equation}
and $|\psi_p\rangle$ is contribution of all other resonances.
\begin{figure}[ht]
\includegraphics[scale=0.4]{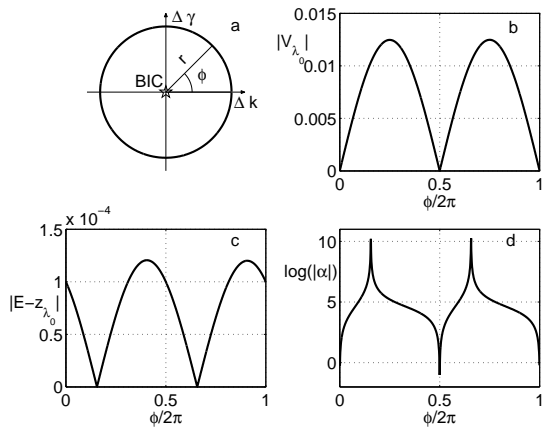}
\caption{Angular behavior of quantities defining the parameter
(\ref{ratio}) around the BIC point marked by star in Fig.
\ref{fig2}.} \label{fig4}
\end{figure}
%\end{center}
%width=.45\textwidth,height=0.25\textheight

As different from the 1d ring, we can study behavior of singular
coefficient (\ref{ratio}) only numerically. Let us encircle the
BIC point $k_0=\sqrt{E_0-\pi^2}, \gamma_0$ as $\Delta k=r\cos\phi,
~\Delta\gamma=r\sin\phi$ as shown in Fig. \ref{fig4} (a) where $r$
is the radius of encircling. Angular behaviors of quantities
defining the parameter $\alpha$ are shown in Fig. \ref{fig4}
(b,c). In particular the numeric shows that $|V_0({\bf s})| \sim
|{\bf s}-{\bf s}_0|^{1/2}$. The behavior of $\alpha$ in Fig.
\ref{fig4} (d) is very similar to the behavior of the parameter
$\alpha=\Delta\gamma/(\Delta k+i\Delta\gamma^2/2)$ for the 1d ring
(see (\ref{psilimit})) except that in the 2d ring we observe phase
difference. As one can see from Fig. \ref{fig4} (d) at $\phi=0,
\pi~ (\Delta\gamma=0)$ the parameter $\alpha=0$, and at
$\phi=\phi_0<\pi/2, ~ \alpha\rightarrow \infty$. The angle
$\phi_0$ exactly corresponds to the direction of tangential line
of transmission zero shown by dashed line in Fig. \ref{fig2}.
Therefore in order to extract the $|\psi_p\rangle$ from the
scattering wave function (\ref{psiR}) we should put at first
$\Delta\gamma=0$ and then limit $\Delta k\rightarrow 0$. If take
limit to the BIC point along  $\Delta\gamma=tan(\phi_0)\Delta k$,
the scattering state transforms to the BIC state $|\lambda_0)$
shown in Fig. \ref{fig3} (a). The particular solution for the
scattering wave function $|\psi_p\rangle$ is shown in Fig.
\ref{fig3} (b).

\section{Conclusion}
Formulas (\ref{BIC sol}) and (\ref{psiRR}) are the key ones which
show that scattering wave function $\overrightarrow{\psi}$ is not
unique since BIC can be superposed with arbitrary coefficient
$\alpha$ to $\overrightarrow{\psi}$. Such kind of decomposition
was established recently for the scattering theory on graphs
\cite{texier1,texier2}. Thus, at the point the ${\bf s}_0$, the
system becomes degenerate. Usual transport solution with energy
$E=E_0$ is complemented by the squared integrable (localized
interior the ring) state $|\lambda_0({\bf s}_0))$ with the same
energy $E_0$ orthogonal to the former. The last state is therefore
BIC. Our consideration shows exactly that BIC is the eigen vector
of the non-hermitian effective Hamiltonian $H_{eff}$ at those
point at which the complex eigenvalue of $H_{eff}$ becomes real
and coincides with the energy of incident particle. The scattering
matrix is unique but not analytical at the BIC points as could be
seen from formula (\ref{Sappr}). As seen from there, the
transmission zeros cross the transmission ones at the BIC point.
Note that these results are not restricted by only AB rings but
applicable for any open quantum system which allow to vary at
least two physical relevant parameters, for example, energy of
incident particles and shape of billiard \cite{sadreev3}.

 BIC is disconnected from both single channel continua. In
order to achieve that BIC is to be a such superposition of eigen
states of closed system that overlapping (\ref{Wlambda}) vanishes
at BIC points. Specifically in the present case of the AB ring
attached to the single channel leads  this superposition becomes
odd function relative to the even function of the leads as seen
from Fig. \ref{fig5} (a). For the 1d ring nodes of BIC are to be
at points of connection of the ring to leads, thereby at those
points where the ratio of lengths of the arms is rational
\cite{texier1}. However for the 2d ring the leads are to be
attached exactly symmetrically as shown in Fig. \ref{fig5}. It
follows then that a violation of symmetry of the system relative
to transport axis $x$ leads to breakdown of BIC. In particular it
occurs for system disordered by impurities. In order BIC could
survive under this violation of symmetry one can use geometry
given in \cite{exner} in which infinite strip attached to the
ring. Moreover impurities lift a degeneracy of closed ring
\cite{chakraborty}. However as shown in \cite{sadreev3} a
condition for BIC to survive is still remaining. From above it
follows that for the system symmetrical relative the y-axis (axis
perpendicular to the transport axis) all odd eigen states of
closed system are BIC's provided that the leads are excited in the
first even channel. Then a perturbation which lifts this symmetry
transforms BIC's into resonance states widths of which are
proportional to the perturbation. The external magnetic field
which subjects only the ring is an example of such a perturbation.

The electron-electron interactions preserve degeneracy of closed
ring \cite{chakraborty}. They modify the energy spectrum and the
coupling between leads and closed ring. As shown in
\cite{sadreev2} variation of the coupling changes a position of
BIC ${\bf s}_0$ however is not important to achieve real value of
the complex eigen value of the effective Hamiltonian. However the
Coulomb interactions might be important in respect that BIC's can
exhibit discrete charging similar to that predicted for resonance
trapping in quantum dots strongly connected to the leads
\cite{berkovits}. The strong coupling of closed quantum system
with leads ($|V_{b,p}^C|\gg |E_b-E_{b'}|$ in terms of
(\ref{Heff})) is hardly achievable while an existence of BICs is
free of a value of the coupling between the closed system and
continua.

Processes of inelastic scattering give rise to finite resonance
width. In that sense BIC is very subtle phenomenon for electron
transmission. However as formulas (\ref{psilimit}) and
(\ref{psiRR}) show, BIC state participates in the scattering wave
function. If above mentioned processes are efficiently small BIC
state can dominate in the vicinity of the BIC point for proper
choice of physical parameters, energy and flux, as shown in Fig.
\ref{fig4} (d).

 {\bf Acknowledgments}. AFS thanks Igor Abrikosov for discussions.
 This work has been
supported by RFBR grant 05-02-97713 "Enisey".

%%%%%%%%%%%%%%%%%%%%%%%%%%%%%%%%%%%%%%%%%%%%%%%%%%%

\end{document}